# Evolution of Magnetic Double Helix and Quantum Criticality near a Dome of Superconductivity in CrAs

M. Matsuda[1=], F. K. Lin[2=], R. Yu[3=], J.-G. Cheng[2,4*], W. Wu[2,4], J. P. Sun[2,4], J. H. Zhang[2,4], P. J. Sun[2,4], K. Matsubayashi[5], T. Miyake[6], T. Kato[5], J.-Q. Yan[1], M. B. Stone[1], Qimiao Si[7], J. L. Luo[2,4,8], and Y. Uwatoko[5]

[1]*Neutron Scattering Division, Oak Ridge National Laboratory, Oak Ridge, Tennessee 37831, USA*

[2]*Beijing National Laboratory for Condensed Matter Physics and Institute of Physics, Chinese Academy of Sciences, Beijing 100190, China*

[3]*Physics Department and Beijing Key Laboratory of Opto-electronic Functional Materials and Micro-nano Devices, Renmin University, Beijing 100872, China*

[4]*School of Physical Sciences, University of Chinese Academy of Sciences, Beijing 100190, China*

[5]*Institute for Solid State Physics, University of Tokyo, 5-1-5 Kashiwanoha, Kashiwa, Chiba 277-8581, Japan*

[6]*Nanosystem Research Institute, AIST, 1-1-1 Umezono, Tsukuba 305-8568, Japan*

[7]*Department of Physics and Astronomy and Rice Center for Quantum Materials, Rice University, Houston, Texas 77005, USA*

[8]*Collaborative Innovation Center of Quantum Matter, Beijing, China*

=These authors contributed equally to this work.
*E-mail: jgcheng@iphy.ac.cn

# Abstract

At ambient pressure CrAs undergoes a first-order transition into a double-helical magnetic state at $T_N$ = 265 K, which is accompanied by a structural transition. The recent discovery of pressure-induced superconductivity in CrAs makes it important to clarify the nature of quantum phase transitions out of its coupled structural/helimagnetic order. Here we show, via neutron diffraction on the single-crystal CrAs under hydrostatic pressure ($P$), that the combined order is suppressed at $P_c \approx$ 10 kbar, near which bulk superconductivity develops with a maximal transition temperature $T_c \approx$ 2 K. We further show that the coupled order is also completely suppressed by phosphorus doping in $CrAs_{1-x}P_x$ at a critical $x_c \approx$ 0.05, above which inelastic neutron scattering evidenced persistent antiferromagnetic correlations, providing a possible link between magnetism and superconductivity. In line with the presence of antiferromagnetic fluctuations near $P_c$ ($x_c$), the $A$ coefficient of the quadratic temperature dependence of resistivity exhibits a dramatic enhancement as $P$ ($x$) approaches $P_c$ ($x_c$), around which $\rho(T)$ has a non-Fermi-liquid form. Accordingly, the electronic specific-heat coefficient of $CrAs_{1-x}P_x$ peaks out around $x_c$. These properties provide clear evidences for quantum criticality, which we interpret as originating from a nearly second-order helimagnetic quantum phase transition that is concomitant with a first-order structural transition. Our findings in CrAs highlight the distinct characteristics of quantum criticality in bad metals, thereby bringing out new insights into the physics of unconventional superconductivity such as occurring in the high-$T_c$ iron pnictides.

The study of unconventional superconductivity (SC) is one of the most vigorous research fields in condensed matter physics.[1] Among the families of unconventional superconductors are the celebrated cuprate[2] and iron-based[3] high-$T_c$ superconductors as well as the heavy-fermion[4] low-$T_c$ ones. Although the underlying mechanisms responsible for the observed SC remains elusive, extensive investigations over the last decades have provided evidences for quantum criticality as a candidate mechanism for these diverse classes of superconductors.[5-7] The phase diagrams of these materials often feature a superconducting dome situated adjacent to a magnetically ordered state, with the optimal superconducting transition temperature ($T_c$) located near a quantum critical point (QCP). The close proximity of SC to magnetic order makes it important to elucidate the nature of magnetism. In the iron-based superconductors, for instance, the mechanism for SC might differ significantly depending on the description of magnetism,[8] which can arise from either the Fermi-surface nesting in the weak coupling limit or the incipient Mott physics when the electron correlations are strong. In addition, the fluctuations associated with the QCP provide a means to understand the non-Fermi-liquid behavior in the normal state. The properties that deviate noticeably from the conventional Fermi-liquid forms include,[7, 9] *e.g.* the resistivity $\rho \propto T^n$ with $1 < n < 2$. In this work, motivated by the recent discovery some of us made of pressure-induced SC in CrAs,[10] we explore the role of magnetic QCP in this system.

CrAs crystallizes in the orthorhombic MnP-type structure at ambient conditions. It is an antiferromagnetic metal and develops a double helical magnetic order below $T_N \approx 265$ K.[11, 12] Interestingly, the antiferromagnetic order at $T_N$ was found to be accompanied by a first-order isostructural transition, which is characterized by a sudden expansion of $b$ axis by ~4% and a slight (< 1%) reduction of $a$ and $c$.[13] Neutron powder diffraction revealed a large ordered Cr moment of ~1.7 $\mu_B$ below $T_N$.[12] The nature of the first-order antiferromagnetic transition in CrAs, in particular, the strong anisotropic lattice changes across $T_N$, have not been well understood. Earlier studies have shown that the coupled structural/helical order can be readily suppressed by either the application of external pressure[14] or chemical substitutions,[15] *e.g.* 5% P for As in CrAs$_{1-x}$P$_x$, providing a model system for studying the quantum criticality associated with the destruction of a helimagnetic order. However, the electrical transport properties for CrAs and related compounds were rarely explored because these samples always develop many cracks or are even shattered into pieces when crossing $T_N$ due to a large volume change. Besides, all earlier studies were performed on the polycrystalline samples. These factors together make the plausible magnetic quantum phase transition unexplored. Work by some of us[16] obtained the first CrAs single crystal and characterized in detail its intrinsic resistivity and magnetic susceptibility. More recently, we found CrAs as the first superconductor among Cr-based compounds.[10] The transport measurements also indicated that the development of SC is accompanied by the reduction and eventual suppression of $T_N$. The findings were later confirmed by Kotegawa *et al.*[17] and their subsequent $^{75}$As-nuclear quadrupole resonance (NQR) study evidenced no coherence effect in the nuclear spin relaxation rate $1/T_1$, thus suggesting an unconventional nature for the observed SC in CrAs.[18]

These observations make CrAs an ideal material for in-depth studies on how the coupled structural and helimagnetic orders are tuned to the critical point and how the magnetism is coupled to the unconventional SC developed nearby. In combinations of elastic and inelastic neutron scattering, resistivity, and specific-heat measurements on undoped CrAs and P-doped $CrAs_{1-x}P_x$ single crystals and powders, we have studied systematically the evolutions of static helical order and dynamic spin correlations as well as the electronic properties of CrAs when its coupled structural/helical order is suppressed by either external pressure or the chemical substitutions. Our results show explicitly the suppression of structural and magnetic transitions by external or chemical pressure, and reveal the persistent antiferromagnetic correlations above $T_N$ and $x_c$ ($P_c$) where the long-range magnetic order is absent. In addition, our results provide clear evidences for magnetic quantum criticality around $x_c = 0.05$ and $P_c \approx 10$ kbar, near which the $T_c$ exhibits a broad maximum. We argue that these results can be understood in terms of a proximity to a magnetic QCP [19] tuned by the degree of the incipient localization of the Cr-3d states,[20] possibly via an orbital-selective Mottness mechanism,[21, 22] which have been proposed for the iron-based high-$T_c$ superconductors.

CrAs and doped $CrAs_{1-x}P_x$ single crystals used in this study were grown out of the Sn flux as described elsewhere.[16] Single-crystal neutron diffraction measurements were performed on a triple-axis spectrometer HB-1 installed at the High Flux Isotope Reactor (HFIR) of Oak Ridge National Laboratory (ORNL). The horizontal collimator sequence was 48'-open-sample-open-240'. The fixed incident neutron energy was 13.5 meV with the energy resolution of ~ 1.5 meV. Since the crystals have needle-like shape with the longest dimension along the $a$ axis, eight pieces of CrAs single crystals were co-aligned in the pressure cell so that the ($0KL$) scattering plane can be measured. The high pressures were generated with a self-clamped piston-cylinder cell made of a Zr-based amorphous alloy[23]. Fluorinert was chosen as the pressure transmitting medium (PTM). The pressure inside the piston-cylinder cell was determined at room temperature by measuring the lattice constant of a comounted NaCl crystal. Cryogenic temperatures down to 4 K were realized by attaching the pressure cell to a closed-cycle $^4$He gas refrigerator. Inelastic neutron scattering (INS) measurements were carried out on the polycrystalline $CrAs_{1-x}P_x$ ($x$ = 0.0 and 0.06) at ambient pressure using the chopper spectrometer SEQUOIA installed at the Spallation Neutron Source (SNS) of ORNL. Measurements of the $dc$ magnetic susceptibility under pressure were performed by using a miniature piston-cylinder cell fitting into the commercial Superconducting Quantum Interference Device (SQUID, Quantum Design). The pressure was monitored by measuring the superconducting transition of Pb. The CrAs crystals were aligned in the pressure cell with $H//a$. Since the magnetic signal from the pressure cell is in the same level as that from CrAs samples, we employed the background record/subtract function in SQUID, which was found to be essential in order to follow correctly the magnetic transition of CrAs under pressure. The resistivity of CrAs single crystal under pressure was measured with the standard four probe method; the pressures up to 21 kbar and 70 kbar were generated by employing a piston-cylinder-type cell[24] and a "Palm" cubic anvil apparatus[25], respectively. In both cases, glycerol was used as PTM. Resistivity and specific heat for the $CrAs_{1-x}P_x$ single crystals at ambient pressure were measured by the Physical Property Measurement System (PPMS, Quantum Design).

Fig. 1 shows the temperature dependence of the lattice constants $b$ and $c$ upon cooling for CrAs single crystal under $P$ = 1 bar, 3.5, 6.9, 9.1, and 12 kbar, and for the P-doped $CrAs_{1-x}P_x$ crystals ($x$ = 0, 0.035, 0.05) at ambient pressure. These lattice constants are calculated from the scattering angle of the nuclear Bragg peaks (020) and (103). The abrupt change in the lattice constants are accompanied with the helimagnetic order. At ambient pressure the $b$ axis is expanded by ~ 3.3% and the $c$ axis is contracted by ~0.8% at $T_N$. The results at ambient pressure are consistent with previous studies on the polycrystalline samples.[12] We measured the temperature dependence of the $b$ and $c$ axes on cooling and heating processes and observed a large hysteresis (see Supplementary Materials Fig. S1), in accordance with the first-order nature of the transition. As shown in Fig. 1(a, b), with increasing pressure the transition temperature decreases and the temperature range where the two phases coexist is enlarged. In addition, the temperature window of the thermal hysteresis also becomes wider with pressure (Fig. S1). At 6.9 kbar, the high-temperature paramagnetic phase is terminated at 70 K upon cooling. At 9.1 kbar, the high-temperature phase is stabilized down to the lowest temperature. However, the low-temperature magnetically ordered phase still exists; its volume fraction is estimated to be ~ 33% at 4 K from the relative intensity of the (020) nuclear Bragg peaks of these two phases (see Fig. S2). Eventually, the structural transition is completely eliminated at 12 kbar. The pressure dependence of the transition temperature for CrAs determined from the present structural study agrees well with that based on the resistivity measurements as shown below.

A characteristic feature of the low-temperature phase below $T_N$ is that both lattice parameters $b$ and $c$ are nearly temperature independent, in contrast to the normal thermal expansion observed at $T > T_N$. In addition, the expansion of $b$ axis when cooling across $T_N$ becomes enhanced with pressure. In contrast for the $c$ axis, the structural anomaly when cooling across $T_N$ changes from a sudden drop at 1 bar to a slight expansion at $P \geq 3.5$ kbar, Fig. 1(b). Notably, this latter observation on the single-crystal sample is different from what observed on the polycrystalline samples that always show a sudden contraction across $T_N$.[26]

The chemical pressure applied via P substitution for As in $CrAs_{1-x}P_x$ plays the same role as the external pressure in suppressing the structural transition of CrAs. As shown in Fig. 1(c, d), both the transition temperature and the temperature dependence of $b$ axis for $x = 0.035$ are similar to those of CrAs under 3.5 kbar, except that the transition is sharper without having a two-phase coexistence region for the P-doped samples. However, the $c$ axis response at $T_N$ differs for the external and chemical pressures: it is expanded slightly under $P = 3.5$ kbar, but contracted suddenly for $x = 0.035$. For the critical composition $x_c = 0.05$, there is no transition down to the lowest temperature, and both $b$ and $c$ decrease gradually with temperature, which are similar to the behaviors of CrAs under 12 kbar. As noted above, the two-phase coexistence region observed under external pressures is absent in the P-doped samples, which exhibit a sharp transition as in the undoped CrAs at ambient pressure. This observation indicates that the external pressure is not applied homogeneously through the whole sample, which might explain the filamentary SC observed in the pressure range 3 < $P$ < 7 kbar.[10] However, superconductivity has not been realized in these P-doped samples presumably because the substantial disorders introduced via chemical doping hinder the Cooper pairing,[10] which will be discussed in detail below.

In order to follow directly the evolution of helimagnetic order for CrAs, we examined a magnetic Bragg peak at (0, 0, 2-$\delta$) as a function of temperature and pressure as well as the P doping. The data are summarized in Fig. 2. It has been well established that the magnetic structure of CrAs at ambient pressure is characterized by two coupled interpenetrating helical sublattices with the propagation vector (0, 0, $\delta$). Instead of new fundamental reflections, such a double helical order of CrAs is featured by some double satellite reflections to the fundamental ones, such as (0, 0, 2±$\delta$). As shown in Fig. 2(a), our results at ambient pressure confirmed the double helical structure with $\delta$ = 0.356 in reciprocal lattice unit (r.l.u.) at 4 K, and that $\delta$ becomes larger with increasing temperature, in excellent agreement with the previous studies on powder samples.[12] With increasing pressure, the magnetic Bragg peak broadens and diminishes gradually with a slight decrease of $\delta$. As shown in Fig. 2(b), no magnetic signal can be observed down to 4 K at $P \geq$ 9.1 kbar. All these results are in general consistent with our previous resistivity measurements under pressure,[10] but provide direct evidence for the disappearance of coupled structural/helical order.

Recently, Keller *et al.*[27] and Shen *et al.*[26] performed high-pressure neutron diffraction experiments on polycrystalline CrAs samples and found evidences in support of the possible connection between magnetism and SC. In particularly, Shen *et al.*[26] reported a pressure-induced spin reorientation from the *ab* to the *ac* plane with an abrupt reduction of $\delta$ at around 6 kbar, above which $\delta$ is well below 0.3. However, our results in Fig. 2(d) shows a $\delta$ = 0.334 at 6.9 kbar, indicating the absence of such transition at least up to 6.9 kbar for the single-crystal sample. Such discrepancy might arise from the different forms of samples studied, *i.e.* single crystal versus polycrystalline samples. As noted above, the lattice constant *c* indeed displays a distinct response to pressure for the single crystal and polycrystalline samples. In addition, a recent high-pressure NQR study on CrAs single crystals also found no clear signature of a change in the magnetic structure under pressure.[18] If we assume that the helical structure does not change other than $\delta$ with pressure and normalize the magnetic signal by the (020) nuclear Bragg intensity, the ordered magnetic moment at 6.9 kbar, which is proportional to the square root of the magnetic intensity, is estimated to be ~ 65% of that at ambient pressure. This estimated moment is between those two polycrystalline results (~25% [26] and ~85% [27]).

In addition to reduce the transition temperature, the P doping also changes the modulation vector of the helical order in a similar manner as the external pressure. Fig. 2(c) shows the magnetic Bragg peak at (0, 0, 2-$\delta$) for $CrAs_{0.965}P_{0.035}$ at different temperatures. For this composition, $\delta$ is 0.342 at 4 K, close to that of CrAs at 3.5 kbar, and $\delta$ increases upon warming up with a larger slope than the undoped CrAs, as shown in Fig. 2(d). No magnetic signal was observed down to 4 K for the $x = 0.05$ sample, in line with the absence of structural transition shown in Fig. 1 (c, d).

From these above comprehensive characterizations based on the elastic neutron scattering, we have demonstrated that the application of external pressure and the P-doping suppress the coupled structural/helimagnetic order in a similar fashion. Thus, we would expect that the structural and magnetic properties should behave similarly for CrAs above $P_c$ and $CrAs_{1-x}P_x$ above $x_c$, where the static order is absent. Such an analogy will allow us to access the spin dynamics of CrAs above $P_c$ via performing inelastic neutron scattering (INS) on the P-doped samples at ambient pressure, because INS under pressure is known to be extremely difficult in view of the tiny sample volume and large background from the pressure cell. Considering the first-order nature of the coupled structural/helimagnetic order, the direct information of dynamic spin correlations above $x_c$ ($P_c$) is crucial for elucidating the interplay between spin fluctuations and SC.

Fig. 3 shows the INS spectra for polycrystalline $CrAs_{1-x}P_x$ ($x = 0$ and 0.06). For CrAs, the sharp and steep magnetic excitations were observed below $T_N$ at $Q = 0.35$, 1.3, 1.8 and 2.15 Å$^{-1}$, which correspond to magnetic Bragg positions at (0, 0, ±$\delta$), (1, 0, 1-$\delta$), (1, 0, 1+$\delta$), and (1, 1, ±$\delta$), respectively. The spectrum at 100 K shown in Fig. 3(a) can illustrate the low-energy magnetic excitations clearly. The shape of the magnetic excitation spectra does not show strong temperature dependence below $T_N$. Only the intensity distribution in energy changes, which follows the Bose factor. Note that there exist phonon scattering above $Q > 2.5$ Å$^{-1}$ and also in the energy ranges of 10 ≤ $E$ ≤ 25 meV and 28 ≤ $E$ ≤ 38 meV. We found that the magnetic excitations at low-$Q$ are extended up to ~110 meV, as shown in Fig. 3(b). Above $T_N$, very broad peaks remain at very low-$Q$ and ~1.6 Å$^{-1}$, Fig. 3(c), indicating that the short-range antiferromagnetic correlations survive above $T_N$, but the steep incommensurate magnetic excitations are broadened. For the nominal CrAs$_{0.94}$P$_{0.06}$ without static magnetic order, a very broad magnetic excitation at $Q$ ~1.0 Å$^{-1}$ was observed. The temperature dependence of $S(|Q|)$, which is integrated with respect to energy between 4.0 and 7.5 meV, is shown in Fig. 3(e). The broad peak at ~1.0 Å$^{-1}$ gradually increases with increasing temperature, which follows the Bose factor. This broad peak feature at 1.0 Å$^{-1}$ becomes less distinct at 95 K, where paramagnetic fluctuations begin to dominate the spectrum. Since $Q$ ~1.0 Å$^{-1}$ corresponds to the antiferromagnetic correlation between next-nearest-neighbor Cr ions, the strongest spin-spin interaction in CrAs, $J_{c2}$, as discussed below, should be the dominant interaction in the paramagnetic state of CrAs$_{0.94}$P$_{0.06}$. Then, short-range antiferromagnetic spin fluctuations due to strong $J_{c2}$ are likely coupled to the SC in CrAs at high pressure. The spin-wave excitation signal from the small amount (~15%) of the long-range ordered state is observable at 50 and 95 K around 1.3 and 1.8 Å$^{-1}$, which are close to those observed in CrAs below $T_N$. In addition, the $T_N$ of the ordered state in CrAs$_{0.94}$P$_{0.06}$ (not shown) is also close to pure CrAs. We thus tend to believe them originating from the phase separated CrAs.

In complementary to the above neutron diffraction study, we also tracked down the pressure dependence of bulk magnetic properties from dc magnetic susceptibility $\chi(T)$ under pressure. Except the $\chi(T)$ curve at 1 bar, all curves shown in Fig. 4 are shifted above for clarity. In agreement with the previous measurement at ambient pressure, the $\chi(T)$ curves at $P$ < 3 kbar increases with temperature and exhibit sudden jump at $T_N$. For $P$ > 3 kbar, the magnetic transition is manifested as a continuous cusp-like anomaly and broadens upwards with increasing pressure. For $P$ > 6 kbar, it is hard to define $T_N$ and, more interestingly, the $\chi(T)$ display a Curie-Weiss-like temperature dependence, which suggests the presence of strong spin fluctuations when the long-range magnetic order gets ready to collapse, in agreement with the above INS results.

Based on the above elastic and inelastic neutron scattering as well as high-pressure magnetic susceptibility measurements, we have not only followed directly the coupled structural/helimagnetic order as a function of external/chemical pressure, but also evidenced persistent short-range antiferromagnetic correlations above $T_N$ or $x_c$ when the static order is absent, thus providing strong evidences in favor of unconventional magnetism mediated SC emerged near a helimagnetic QCP.

To search for further evidences of quantum criticality, we have performed high-pressure resistivity measurements on CrAs single crystal up to 70 kbar and the resistivity and specific-heat measurements on $CrAs_{1-x}P_x$ single crystals at ambient-pressure. The low-temperature resistivity data of CrAs under various pressures up to 70 kbar are shown in Fig. 5(a, b). The present high-pressure resistivity study considerably extends the pressure range of previous studies,[10],[17] allowing us to construct a complete superconducting phase diagram shown in Fig. 5(f), and also to characterize in detail the helimagnetic quantum criticality. As shown in Fig. 5(b), the superconducting transition temperature $T_c$ decreases gradually and reaches below 0.5 K at $P > 35$ kbar, resulting in a dome-shaped $T_c(P)$ with a broad maximum of $T_c$ around $P_c \approx 10$ kbar, Fig. 5(f). The normal-state $\rho(T)$ under pressures up to 70 kbar are displayed in Fig. 5(c, d) in the form of $\Delta\rho(=\rho-\rho_0)$ versus $T^2$. As can be seen, $\Delta\rho(T^2)$ under $P < 3$ kbar follow nicely the linear behavior below 10 K. A clear deviation from the linear dependence is evidenced for $P > 3$ kbar. The strongest deviations occur in the intermediate pressure range $5 < P < 13$ kbar. The sublinear behavior in $\Delta\rho$ versus $T^2$ plot indicated a resistivity exponent $n < 2$ in $\rho \sim T^n$. With increasing $P > 20$ kbar, the linear dependence is gradually restored. We thus have fitted the resistivity data below 10 K with the general power-law formula: $\rho(T) = \rho_0 + BT^n$, as shown by the broken lines in Fig. 5(a), and extracted the exponent $n$, which contains valuable information about the nature of the metallic state. The Landau Fermi-liquid (FL) theory predicts an $n = 2$ for correlated metals, while the observation of $n < 2$ has been taken as a common signature for non-Fermi-liquid (nFL) metals near a QCP. As shown in Fig. 5(e), the application of high pressure that suppresses $T_N$ also converted the behavior from FL to nFL; $n$ deceases quickly to a minimum of 1.3(1) at ~ 6 kbar, slightly lower than $P_c$. For $P > 6$ kbar, the exponent $n$ increases gradually and reaches ~1.82 at 70 kbar. Although the above analysis evidenced a dramatic enhancement of the $B$ coefficient around $P_c$, the variation of exponent $n$ prevents a quantitative comparison. We therefore perform a linear fit to the $\Delta\rho$ versus $T^2$ plot in the low-temperature limit, Fig. 5(c, d), which yields the $T$-quadratic coefficient $A$ defined by $\rho-\rho_0 = AT^2$. As shown in Fig. 5(g), the $A$ coefficient displays a pronounced peak centered at $P_c$. Since the $A$ coefficient of FL is related to the effective mass of charge carriers via $A \propto (m^*/m_0)^2$, the significant enhancement of $A$, especially on the side of $P < P_c$, signifies a dramatic enhancement of the effective mass associated with the suppression of helimagnetic order. The observations of nFL behavior and dramatic enhancement of the $A$ coefficient for CrAs near $P_c$ thus provide important clues for the presence of magnetic quantum criticality associated with the helimagnetic order.[28, 29] This point is further supported by the P-doped $CrAs_{1-x}P_x$ samples.

Fig. 6 summarizes the resistivity and specific-heat results on the $CrAs_{1-x}P_x$ samples. As can be seen in Fig. 6(a), the first-order structural/helical order of CrAs manifested as a sudden change in resistivity is gradually suppressed to lower temperatures by P doping, and disappears completely in the $x = 0.05$ sample, in agreement with the structural data shown in Fig. 1(c, d). It should be noted that the resistivity curves for the $0 < x < 0.05$ samples exhibit a strong jump, instead of a drop as seen in CrAs, at $T_N$ upon cooling. As a result, all these P-doped samples are featured by a large residual resistivity $\rho_0$ in the range of 40-70 $\mu\Omega$ cm, which should be attributed to presence of chemical disorders and/or strains induced by the large volume change at the structural transition. We have shown in our previous study[10] that the pressure-induced SC in CrAs is very sensitive to the residual resistivity $\rho_0$, and bulk SC can only be achieved on high-quality CrAs single crystals with $\rho_0 \approx 1\text{-}2$ $\mu\Omega$ cm, in which the electron mean free path $l_{mfp}$ is large than the superconducting coherent length $\xi$. Such a criterion, however, cannot be fulfilled for these P-doped samples having a very large $\rho_0$, which can rationalize the absence of SC for these P-doped samples. Nonetheless, characteristics of QCP are observed near $x_c$. As seen in Figs. 6(b, c, d), nFL behavior with a resistivity exponent $n \approx 1.5$ and a dramatic enhancement of the $T$-quadratic coefficient $A$ were clearly evidenced at the critical $x_c$ from the analysis of low-temperature resistivity data. Moreover, the advantage of accessing the magnetic QCP at ambient pressure for these P-doped samples is that it allows us to probe directly the electronic properties near the QCP via the specific heat $C(T)$ measurements. The low-temperature $C(T)$ of $CrAs_{1-x}P_x$ single crystals are plotted in Fig. 6(e) in the form of $C/T$ vs $T^2$. A linear fit has been applied on these data to extract the electronic specific-heat coefficient $\gamma$, which was plotted in Fig. 6(g) together with the $A$ coefficient. As can be seen clearly, both quantities exhibit significant enhancement near $x_c$, thus providing unequivocal proof for the presence of magnetic QCP in CrAs.

In order to understand how quantum criticality can arise from the coupled magnetic/structural transition, we carry out an analysis within the framework of Ref. [20]. CrAs has a large room-temperature resistivity,[16] which reaches the Ioffe-Regel limit. It is therefore a bad metal, just as the parent iron pnictides[30]. Consequently, its single-electronic spectrum can be divided into an incoherent part responsible for the quasi-localized moment and a coherent part near the Fermi energy for metallic conduction.[21] A magnetic quantum phase transition may then arise when the spectral weight of the coherent part increases to certain threshold at the expense of the incoherent part.[20] To make further progress, we model the exchange interactions in a Heisenberg form, $H = \sum_{i,j} J_{ij} \mathbf{S}_i \cdot \mathbf{S}_j$, where $J_{ij}$ contains the leading exchange couplings up to fourth nearest neighbors[31] denoted as $J_a$, $J_b$, $J_{c1}$, and $J_{c2}$, as illustrated in Fig.7(a). The ground-state phase diagram is shown in Fig. 7(b). The double-helical magnetic order is stabilized between the two hyperbola in the second and fourth quadrants. With the helical wavevector $\delta = 0.353$, the phase angle for the double-helical order $\alpha \sim \pi/3$ arises when $J_{c1}/J_a \sim -0.52$ and $J_{c2}/J_a \sim 7.01$, as shown in Fig. 7(c), indicating a strong bond overlap along the $c$ axis. Because $J_{c2}$ is considerably stronger than $J_{c1}$ and $J_a$, the band maximum of the magnetic excitations at low-$Q$ that are extended up to ~110 meV in the INS of CrAs shown in Fig. 3(b) could be roughly expressed as $2SJ_{c2}$, where $S$ is the spin value. Since the effective $S \sim 0.85$, $J_{c2}$ is estimated to be ~65 meV.

Both $\delta$ and $\alpha$ vary slightly in the magnetic phase, which suggests that the exchange couplings are sensitive to lattice distortions. We model this through a Ginzburg-Landau free energy function:

$$f = a\Delta + \frac{b}{2}\Delta^2 - \frac{v}{2}\Delta m^2 + \frac{r}{2}m^2 + \frac{u}{4}m^4,$$

where $m$ and $\Delta$ respectively refer to the helimagnetic order parameter and the unit-cell volume change. It follows from this function that the finite-temperature magnetic and structural transitions are either both first-order or both second-order. This leads to two possible types of the phase diagram for the pressure tuning of the coupled magnetic/structural transitions, as shown in Fig. 7(d). The zero-temperature magnetic transition is treated by a generalization of the Ginzburg-Landau function to an action, in which the order parameters depend on both wavevector and frequency. A damping term for the magnetic order parameter occurs in this action, reflecting the coupling of the local moments to the coherent itinerant electrons. This term makes it possible for the magnetic quantum phase transition to be nearly second order[20], thereby yielding a large dynamical range for magnetic quantum criticality. In CrAs, the nearly continuous nature of the magnetic transition at $T = 0$ is reflected in the gradual reduction of the jump in the dc magnetic susceptibility at $T_N$ as pressure is increased, Fig. 4. The structural transition, however, can remain to be first order. This is consistent with the prominent persistence under pressure of the jump in the lattice constant and the thermal hysteresis reported here, as well as the apparent first-order behavior observed in the NQR measurements.[18]

Our analysis encodes the bad-metal characteristics of CrAs. To further highlight this, it is instructive to compare CrAs with Cr, a good metal with a room-temperature resistivity far below the Ioffe-Regel limit. None of the characteristics of the quantum criticality discussed above for CrAs -- *viz.* the divergent tendency of both the resistivity *A* coefficient and the effective carrier mass, the nFL form of the resistivity, and the emergence of SC in its vicinity -- occurs near the magnetic quantum phase transition in Cr under either hydrostatic or chemical pressure.[32-34] By contrast, these same features have been observed in P-doped iron arsenides[29, 35], which are also bad metals with strong correlations. Our results in CrAs, therefore, highlight the distinct features of quantum criticality in bad metals, including its ability to drive SC. This provides new insights into the essential role the bad metal behavior plays in the emergence of unconventional SC, which appears also important in other correlated superconductors such as the iron pnictides.[36]

The bad metal behavior in CrAs indicates that the electron correlations play a crucial role in this material. Given the multiorbital character of the compound, an important ingredient responsible for our results is the orbital-selective Mottness (OSM), which has been actively discussed recently in the iron pnictides. The OSM in the case of CrAs is characterized by three prominent features: (i) it remains metallic in the whole temperature range, (ii) it acquires below $T_N$ a considerably large localized magnetic moment 1.7 $\mu_B$, which is yet smaller than the fully localized value of $3\mu_B$ for $Cr^{3+}$, (iii) its lattice constants undergo a strong anisotropic change at $T_N$ with the abrupt expansion of the *b*-axis upon cooling below $T_N$. These observations can be rationalized by considering the distinct localized/itinerant characters of Cr-3d orbitals, which are split into doubly degenerate $e_g$ and triply degenerate $t_{2g}$ states in an octahedral crystal field. Since the $e_g$ orbitals are not filled for $Cr^{3+}$, the $t_{2g}$ orbitals ($d_{xy}$, $d_{yz}$, and $d_{zx}$) are mainly involved for the observed electrical and magnetic properties. As discussed previously,[11, 14] $d_{xy}$ orbitals that are directed along the *a* axis with the shortest Cr-Cr nearest-neighbor distance should form a broad band below the Fermi level, $E_F$, while the $d_{yz}$ and $d_{zx}$ orbitals form narrow bands near the Fermi level and their bandwidths and positions vary sensitively depending on change of lattice constants. First-principles calculations[37] on the series of CrX (X = P, As, Sb) indeed found a systematic reduction of bandwidth near $E_F$ (or enhancement of magnetism) with increasing the *b* axis (the size of X anion). This is also consistent with the observation of sudden expansion of *b* axis below $T_N$ that gives rise to a large localized moment of 1.7 $\mu_B$ in CrAs. It is such an intimate correlation between structure and orbital-dependent electronic states that enables pressure to be a very effective knob to tune the system towards the electronic instability and in favor of superconductivity. Such an orbital dependent correlation effect is quite similar with the iron pnictides.[21, 22]

In summary, we have performed a comprehensive elastic and inelastic neutron scattering study on the CrAs under hydrostatic pressures up to 12 kbar and under chemical pressure via P substitutions for As. Our results provide evidence for quantum criticality associated with the double helical magnetic order, and its association with a diverging carrier mass and an emergent superconductivity.

## Acknowledgements

Work at IOP/CAS was supported by the National Science Foundation of China (Grant Nos. 11574377, 11025422, 11674375, 11634015), the National Basic Research Program of China (Grant Nos. 2018YFA0305700, 2014CB921500, 2017YFA0302901), the Strategic Priority Research Program and Key Research Program of Frontier Sciences of the Chinese Academy of Sciences (Grant Nos. XDB07020100, XDB01020300, QYZDB-SSW-SLH013). Work at ISSP/UT was partially supported by Grant-in-Aid for Scientific Research, KAKENHI (Grant Nos. 23340101, 252460135), and the JSPS fellowship for foreign researchers (Grant No. 12F02023). R.Y. was partially supported by the National Science Foundation of China (Grant No. 11374361), and the Fundamental Research Funds for the Central Universities and the Research Funds of Remnin University of China. Work at Rice Univ. was supported by the U.S.

Department of Energy, Office of Science, Basic Energy Sciences, under Award No. DE-SC0018197, and the Robert A. Welch Foundation Grant No. C-1411. Research conducted at ORNL's High Flux Isotope Reactor and Spallation Neutron Source are sponsored by the Scientific User Facilities Division, Office of Basic Energy Sciences, US Department of Energy. This study was supported in part by the U.S.-Japan Cooperative Program on Neutron Scattering. We are grateful to Drs. Kazuki Komatsu (Univ. of Tokyo) and Yoshihiko Yokoyama (IMR, Tohoku Univ.) for use of the Zr-based amorphous pressure cell.

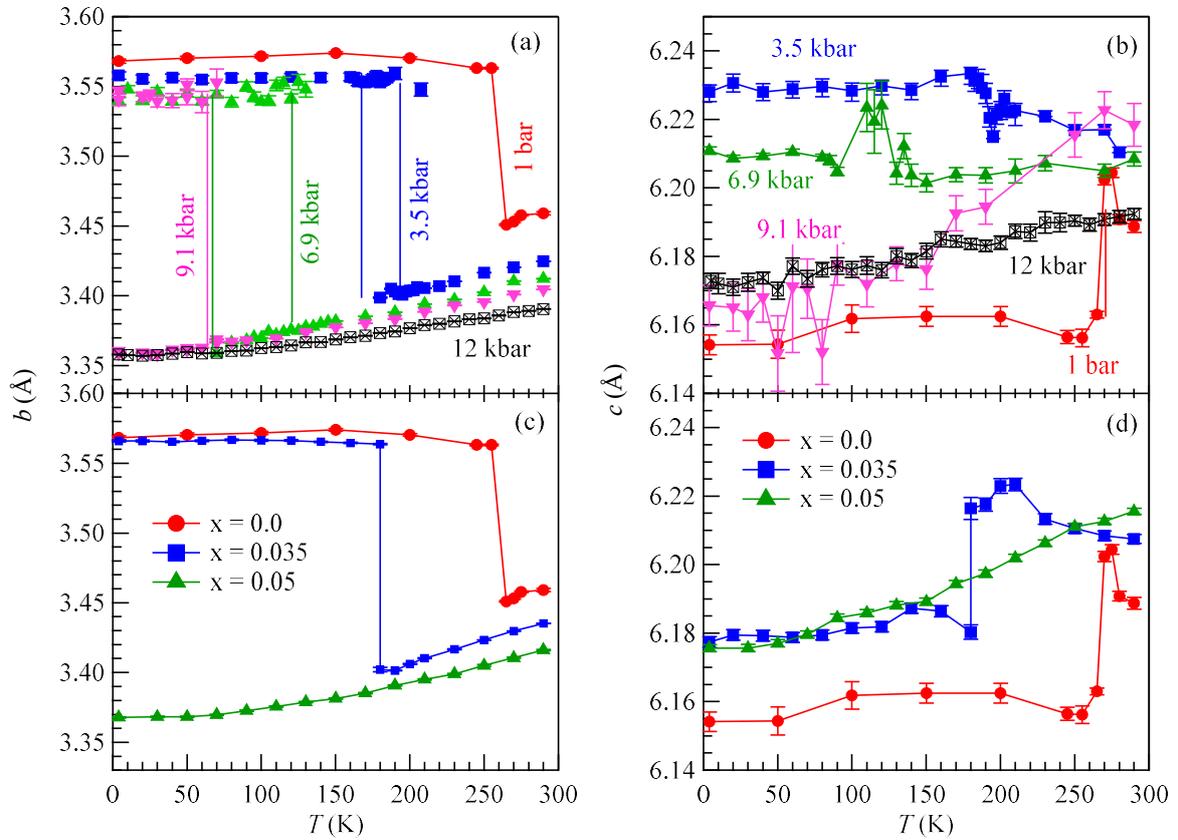

Fig. 1(Color online) Temperature dependence of the lattice constants $b$ and $c$ for CrAs under (a, b) external pressures up to 12 kbar and (c, d) chemical pressure via P substitution for As in $CrAs_{1-x}P_x$ ($x$ = 0, 0.035, 0.05). These data were obtained during the cooling process, in which the temperature ranges marked by the two solid lines in (a) represent the two-phase coexistence region.

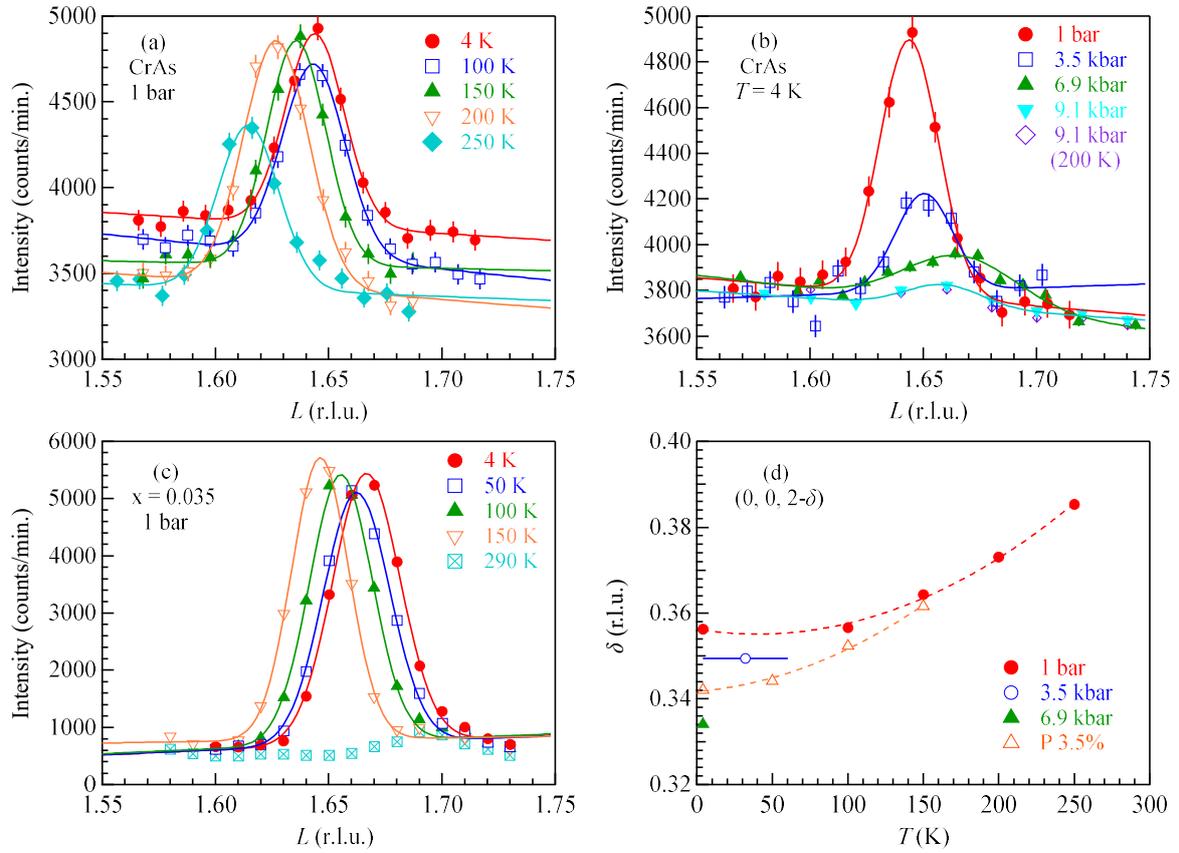

Fig. 2(Color online) The (0, 0, 2-$\delta$) magnetic Bragg peak for CrAs (a) at different temperatures under 1 bar and (b) under different pressures at 4 K, and (c) for 3.5%-P doped CrAs at different temperatures under 1 bar. (d) A summary of $\delta$ as a function of temperature and pressure for undoped and 3.5%-P doped CrAs.

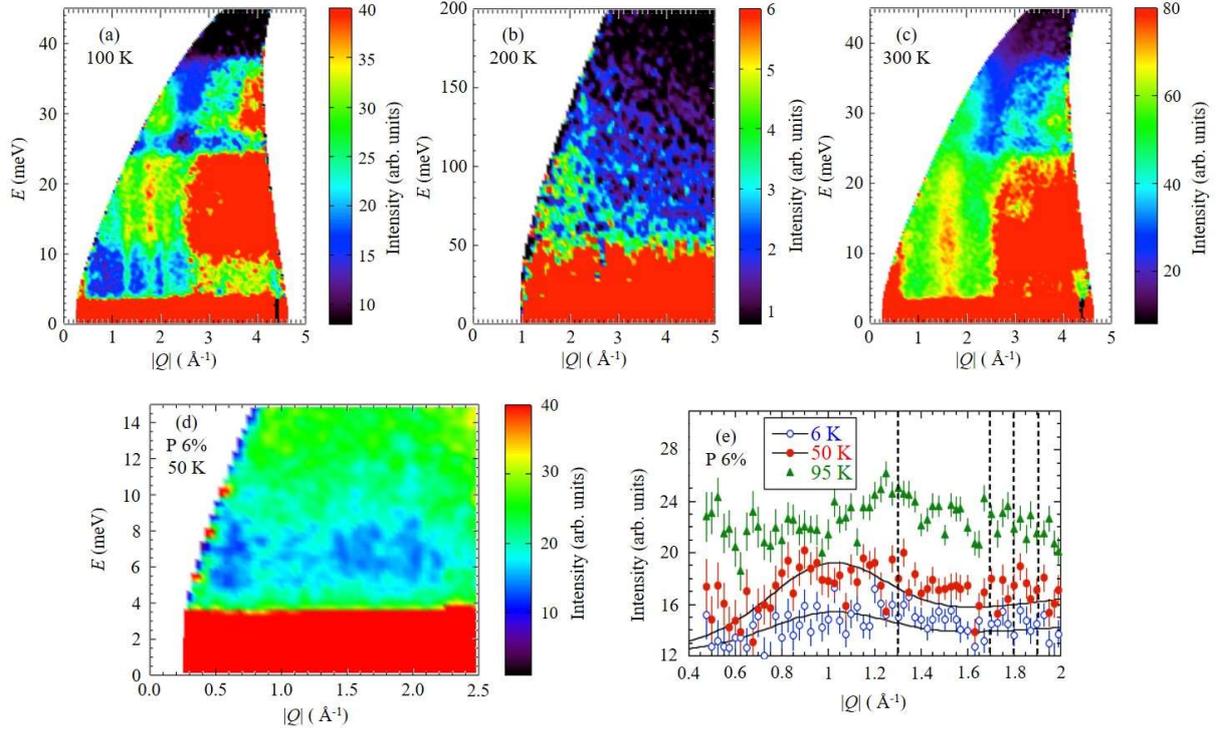

Fig. 3(Color online) Color contour maps of the INS intensity $S(|Q|, E)$ for polycrystalline CrAs measured at (a) $T = 100$ K and (c) 300 K with $E_i = 50$ meV and at (b) $T = 200$ K with $E_i = 800$ meV. Background scattering measured with an empty sample can is subtracted. (d) $S(|Q|, E)$ for polycrystalline CrAs$_{0.94}$P$_{0.06}$ measured with $E_i = 50$ meV at $T = 50$ K. The intensity scales of the two samples are normalized using incoherent scattering. (e) $S(|Q|)$ integrated with respect to energy between 4.0 and 7.5 meV at $T = 6$, 50, and 95 K for CrAs$_{0.94}$P$_{0.06}$. The solid lines are guides to the eye. The broken lines represent the positions where spin-wave excitations from the magnetically ordered phase are expected.

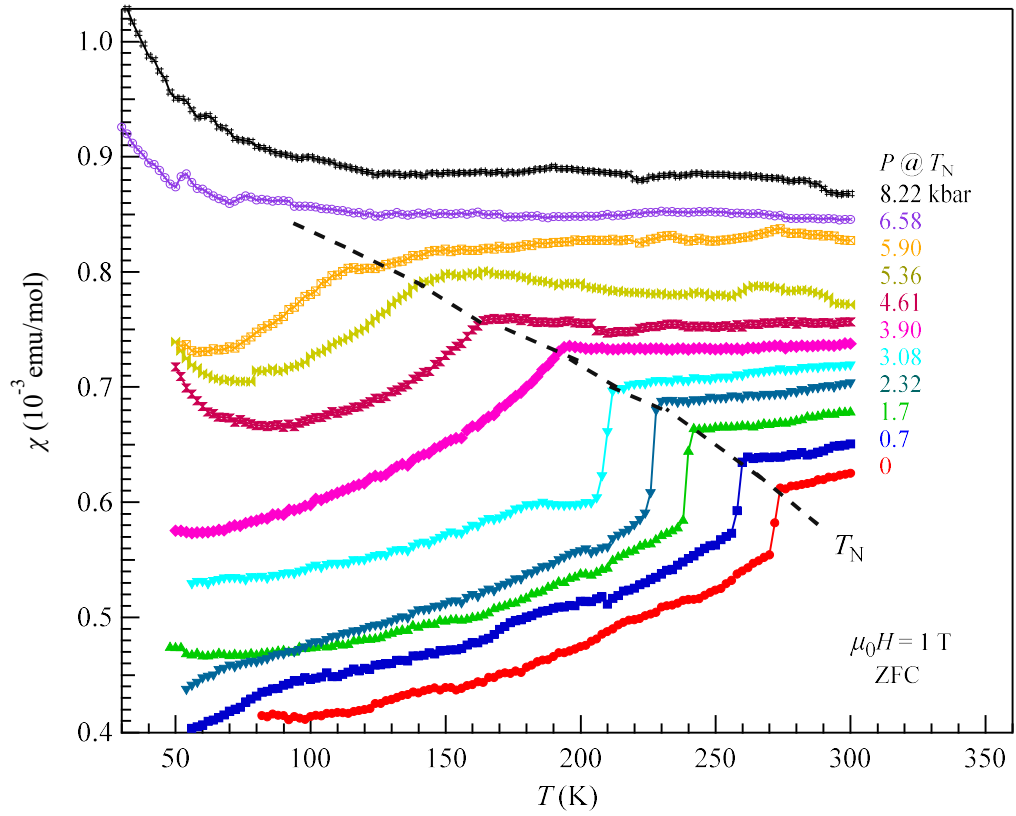

Fig. 4(Color online) Temperature dependence of the *dc* magnetic susceptibility $\chi(T)$ of CrAs single crystals under various hydrostatic pressures up to 8 kbar. Except the $\chi(T)$ curve at 1 bar, all curves have been shifted above for clarity.

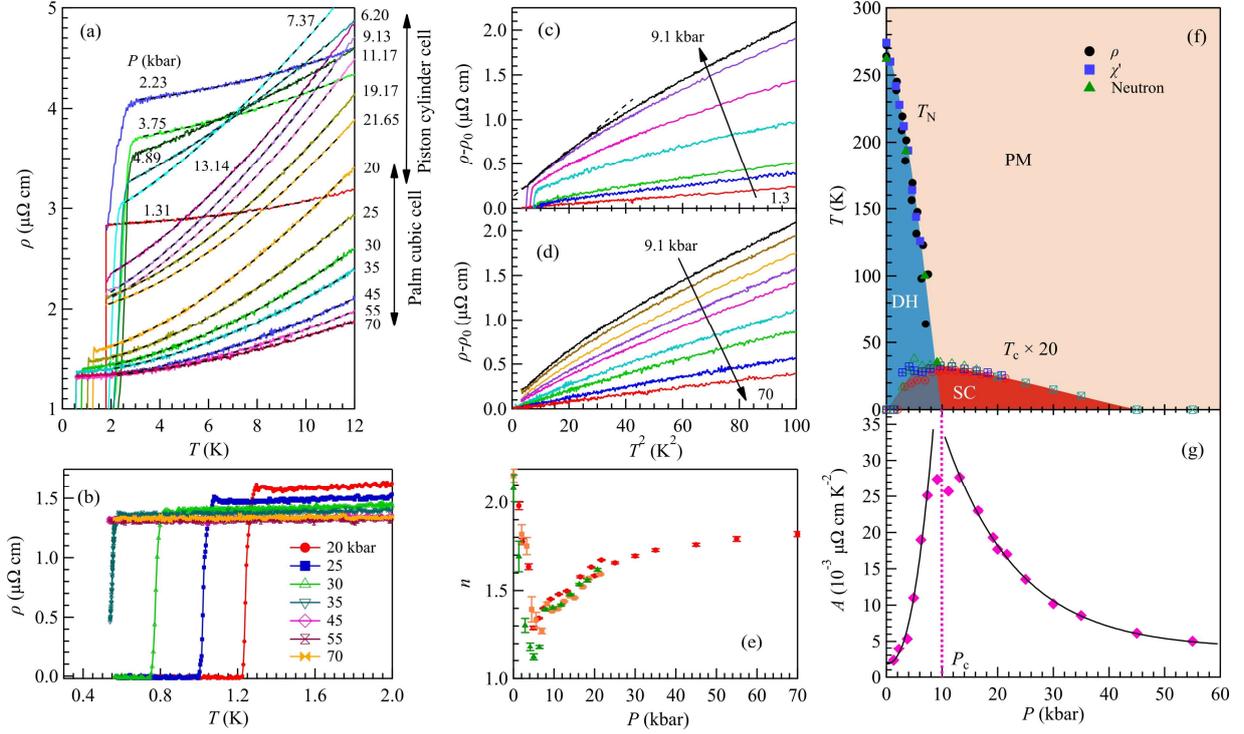

Fig. 5(Color online) (a) Temperature dependence of the resistivity $\rho(T)$ for CrAs single crystal under various pressures up to 70 kbar measured in a piston cylinder cell and a palm cubic anvil cell as indicated in the figure. The $\rho(T)$ data were fitted with the power-law: $\rho(T) = \rho_0 + BT^n$ and the fitting curves are shown as the broken lines. The obtained exponent $n$ as a function of pressure is given in (e). The square and triangle symbols represent the results from the other two samples. The $\rho(T)$ data below 2 K in the pressure range 20-70 kbar are shown in (b) to illustrate the gradual suppression of $T_c$ to below 0.5 K. (c, d) Plots of $(\rho-\rho_0)$ vs $T^2$ for CrAs under various pressures highlighting the evolution of the quadratic coefficient $A$ and deviation from Fermi-liquid behavior. A linear fitting was applied in the low-temperature region to extract the coefficient $A$, which is displayed in (g). (f) An updated $T$-$P$ phase diagram of CrAs showing the complete suppression of $T_c$ around 45 kbar. DH denotes for double helical order, SC for superconductivity, PM for paramagnetism. The coefficient $A$ for CrAs displays an apparent enhancement around $P_c \approx 10$ kbar in (g).

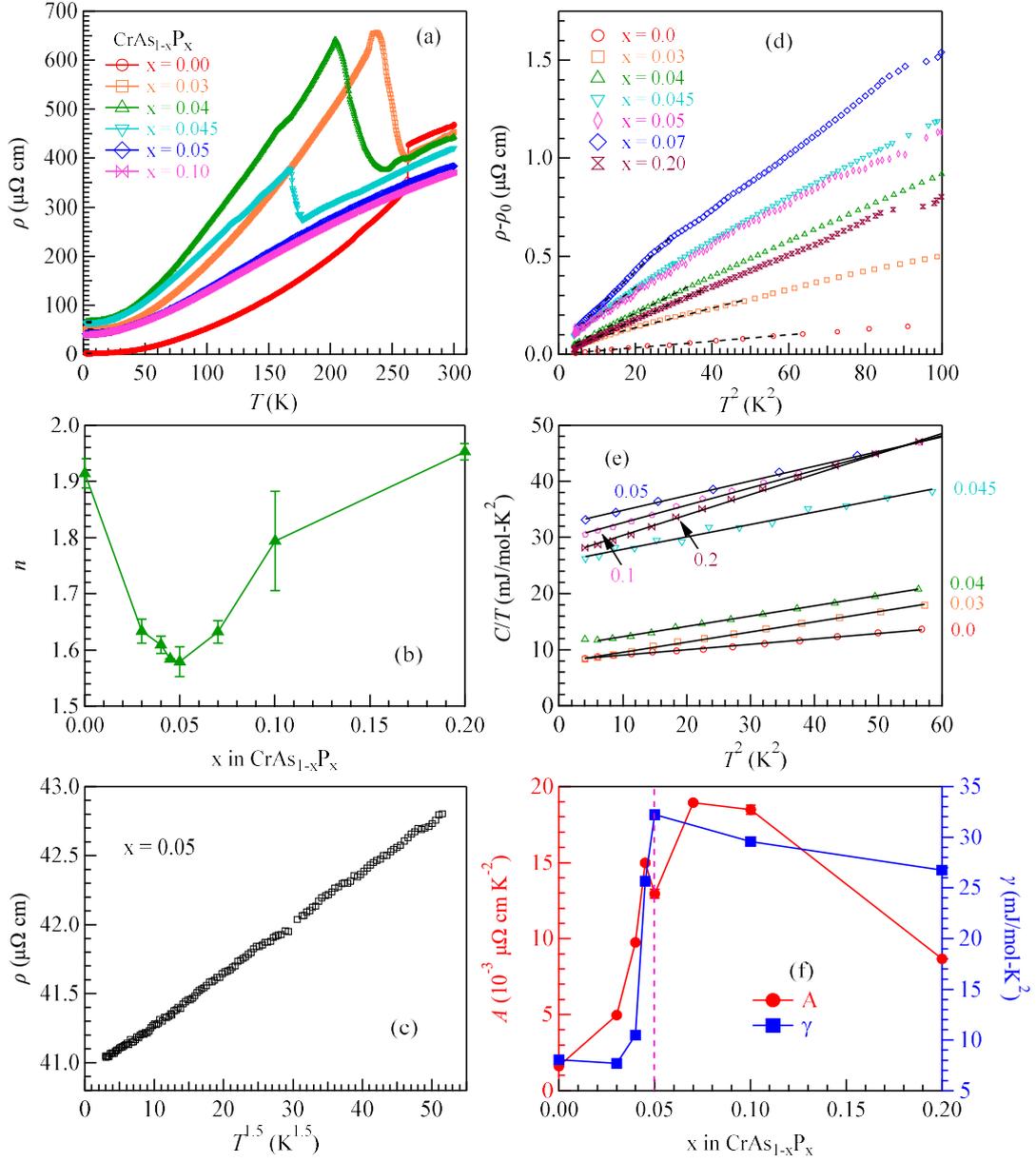

Fig. 6(Color online) (a) Temperature dependence of resistivity $\rho(T)$ for CrAs$_{1-x}$P$_x$ single crystals. The $\rho(T)$ curves below 12 K were fitting to $\rho(T) = \rho_0 + BT^n$ and the obtained exponent $n$ as a function of $x$ is given in (b). (c) A plot of $\rho$ vs $T^{1.5}$ for $x = 0.05$ sample illustrating a nice linear (nFL) behavior. (d) ($\rho$-$\rho_0$) vs $T^2$ and (e) $C/T$ vs $T^2$ plots for CrAs$_{1-x}$P$_x$ samples, from which the $T$-quadratic resistivity coefficient $A$ and the electronic specific-heat coefficient $\gamma$ were extracted, and displayed in (f) as a function of $x$.

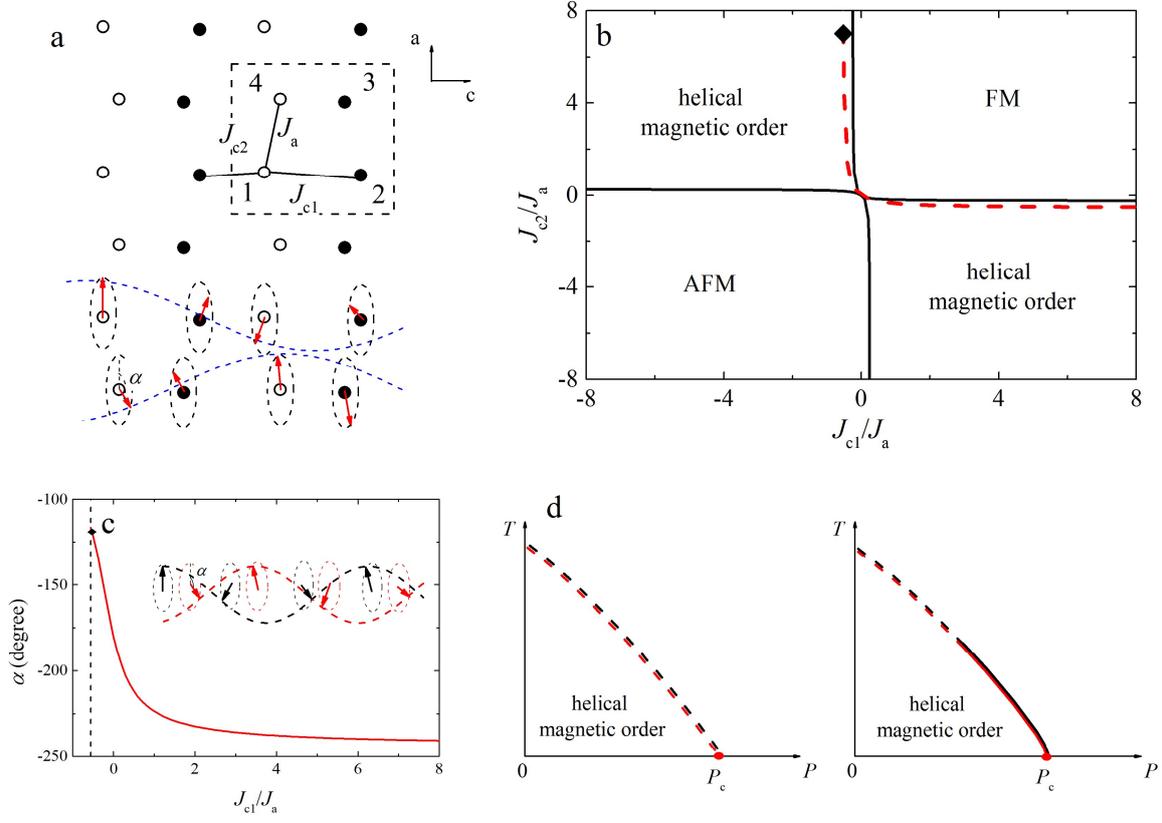

Fig. 7 (Color online) Structure and magnetic phase diagram of CrAs. (a) Sketch of the CrAs lattice, with the *b*-axis perpendicular to the paper plane. The open and closed symbols stand for the Cr atoms at layers $b/4$ and $3b/4$, respectively. Each unit cell consists of four Cr atoms, labeled as 1 through 4. The magnetic ground state is studied in a Heisenberg model[31] with exchange couplings up to fourth nearest neighbor pairs of Cr, $J_a$, $J_b$, $J_{c1}$, and $J_{c2}$, where $J_b$ (not shown) couples the magnetic moments of the same sublattice in nearest-neighboring unit cells along the *b* axis. The arrows label the spin pattern associated with the double helical magnetic order. (b) Ground state magnetic phase diagram of the model for CrAs. FM and AFM refer to ferromagnetic and antiferromagnetic states, respectively. The AFM state has an ordered wavevector $\boldsymbol{Q} = (0, 0, 0)$, but the spins are staggered within each unit cell. The double helical magnetic order has a propagating wavevector $\boldsymbol{Q} = (0, 0, \delta)$. The phase boundaries among the three phases are separated by two hyperbola, marked by solid curves. The red dashed line is a curve with fixed helical wavevector $\delta = 0.353 \cdot 2\pi/c$, and the diamond shows the position of the state with magnetic phase angle $\alpha = -120°$. (c) Evolution of the phase angle $\alpha$ for the double-helical order (illustrated in the inset) with $J_{c1}/J_a$ along the fixed $\delta = 0.353 \cdot 2\pi/c$ curve in panel (b).

(d) Two possible scenarios of the structural (black curve) and magnetic (red curve) phase transitions in CrAs under pressure. The dashed and solid curves refer to first- and second-order transitions, respectively. The structural and magnetic transitions are always coupled together. They are first order at low pressures. With increasing pressure, they can either remain first order (left panel), or become second order (right panel).

# References

1. M. R. Norman, *The Challenge of Unconventional Superconductivity*, Science **332**, 196 (2011).
2. J. G. Bednorz and K. A. Muller, *Possible High $T_c$ Superconductivity in the Ba-La-Cu-O System*, Z. Phys. B **64**, 189 (1986).
3. Y. Kamihara, T. Watanabe, M. Hirano, and H. Hosono, *Iron-Based Layered Superconductor $La[O_{1-x}F_x]FeAs$ (x = 0.05-0.12) with $T_c$ = 26 K*, J. Am. Chem. Soc. **130**, 3296 (2008).
4. F. Steglich, J. Aarts, C. D. Bredl, W. Lieke, D. Meschede, W. Franz, and H. Schafer, *Superconductivity in the Presence of Strong Pauli Paramagnetism: $CeCu_2Si_2$*, Phys. Rev. Lett. **43**, 1892 (1979).
5. P. Coleman and A. J. Schofield, *Quantum Criticality*, Nature **433**, 226 (2005).
6. S. Sachdev and B. Keimer, *Quantum Criticality*, Phys. Today **64**, 29 (2011).
7. P. Gegenwart, Q. Si, and F. Steglich, *Quantum Criticality in Heavy Fermion Metals*, Nature Phys. **4**, 186 (2008).
8. P. Dai, J. Hu, and E. Dagotto, *Magnetism and Its Microscopic Origin in Iron-Based High-Temperature Superconductors*, Nature Phys. **8**, 709 (2012).
9. H. v. Lohneysen, A. Rosch, M. Vojta, and P. Wolfle, *Fermi-Liquid Instabilities at Magentic Quantum Phase Transition*, Rev. Mod. Phys. **79**, 1015 (2007).
10. W. Wu, J.-G. Cheng, K. Matsubayashi, P. P. Kong, F. K. Lin, C. Q. Jin, N. L. Wang, Y. Uwatoko, and J. L. Luo, *Superconductivity in the Vincinity of Antiferromagnetic Order in CrAs*, Nat. Comm. **5**, 5508 (2014).
11. H. Boller and A. Kallel, *First Order Crystallographic and Magnetic Phase Transition in CrAs*, Solid State Comm. **9**, 1699 (1971).
12. K. Selte, A. Kjekshus, W. E. Jamison, A. F. Andresen, and J. E. Engebretsen, *Magnetic Structure and Properties of CrAs*, Acta Chem. Scand. **25**, 1703 (1971).
13. N. Kazama and H. Watanabe, *Magnetic Properties of $Cr_{1-x}Mn_xAs$ System*, J. Phys. Soc. Jpn. **30**, 1319 (1971).
14. E. A. Zavadskii and I. A. Sibarova, *Some Features of Phase Transition in Chromium Arsenide at High Pressures*, Sov. Phys. JETP **51**, 542 (1980).
15. K. Motizuki, H. Ido, T. Itoh, and M. Morifuji, in *Springer Series in Materials Science* edited by R. Hull, C. Jagadish, R. M. J. Osgood, J. Parisi, Z. Wang and H. Warlimont, (2009), Vol. 131.
16. W. Wu, X. D. Zhang, Z. H. Yin, P. Zheng, N. L. Wang, and J. L. Luo, *Low Temperature Properties of Pnictide CrAs Single Crystal*, Science China **53**, 1207 (2010).
17. H. Kotegawa, S. Nakahara, H. Tou, and H. Sugawara, *Superconductivity of 2.2 K under Pressures in Helimagnet CrAs*, J. Phys. Soc. Jpn. **83**, 093702 (2014).
18. H. Kotegawa, S. Nakahara, R. Akamatsu, H. Tou, H. Sugawara, and H. Harima, *Detection of an Unconventional Superconducting Phase in the Vicinity of the Strong First-Order Magnetic Transition in CrAs Using $^{75}As$-Nuclear Quadrupole Resonance*, Phys. Rev. Lett. **114**, 117002 (2015).

19 F. Hammerath, P. Bonfa, S. Sanna, G. Parndo, R. De Renzi, Y. Kobayashi, M. Sato, and P. Carretta, *Poisoning Effect of Mn in LaFe$_{1-x}$Mn$_x$AsO$_{0.89}$F$_{0.11}$: Unveiling a Quantum Critical Point in the Phase Diagram of Iron-Based Superconductors*, Phys. Rev. B **89**, 134503 (2014).

20 J. Dai, Q. Si, J. X. Zhu, and E. Abrahams, *Iron Pnictides as a New Setting for Quantum Criticality*, PNAS **106**, 4118 (2009).

21 Z. P. Yin, K. Haule, and G. Kotliar, *Magnetism and Charge Dynamics in Iron Pnictides*, Nat. Phys. **7**, 294 (2011).

22 L. de' Medic, G. Giovannetti, and M. Capone, *Selective Mott Physics as a Key to Iron Superconductors*, Phys. Rev. Lett. **112**, 177001 (2014).

23 K. Komatsu, K. Mnuakata, K. Matsubayashi, Y. Uwatoko, Y. Yokoyama, K. Sugiyama, and M. Matsuda, *Zr-Based Bulk Metallic Galss as a Cylinder Material for High Pressure Apparatus*, High Pressure Research **35**, 254 (2015).

24 Y. Uwatoko, *Design of a Piston Cylinder Type High-Pressure Cell*, Rev. High Press. Sci. Tech. **12**, 306 (2002).

25 Y. Uwatoko, K. Matsubayashi, K. Matsubayashi, N. Aso, M. Nishi, T. Fujiwara, M. Hedo, S. Tabata, K. Takagi, M. Tado, and H. Kagi, *Development of Palm Cubic Anvil Apparatus for Low Temperature Physics*, Rev. High Press. Sci. Tech. **18**, 230 (2008).

26 Y. Shen, Q. Wang, Y. Hao, B. Pan, Y. Feng, Q. Huang, L. W. Harriger, J. B. Leao, Y. Zhao, R. M. Chisnell, J. W. Lynn, H. B. Cao, J. P. Hu, and J. Zhao, *Structure and Magnetic Phase Diagram of CrAs and Its Relationship with Pressure-Induced Superconductivity*, Phys. Rev. B **93**, 060503(R) (2016).

27 L. Keller, J. S. White, M. Frontzek, P. Babkevich, M. A. Susner, Z. C. Sims, A. S. Sefat, H. M. Ronnow, and C. Ruegg, *Pressure Dependence of the Magnetic Order in CrAs: A Neutron Diffraction Investigation*, Phys. Rev. B **91**, 020409(R) (2015).

28 N. D. Mathur, F. M. Grosche, S. R. Julian, I. R. Walker, D. M. Freye, R. K. W. Haselwimmer, and G. G. Lonzarich, *Magnetically Mediated Superconductivity in Heavy Fermion Compounds*, Nature **394**, 39 (1998).

29 S. Kasahara, T. Shibauchi, K. Hashimoto, K. Ikada, S. Tonegawa, R. Okazaki, H. Shishido, H. Ikeda, H. Takeya, K. Kirata, T. Terashima, and Y. Matsuda, *Evolution from Non-Fermi- to Fermi-Liquid Transport Via Isovalent Doping in BaFe$_2$(As$_{1-x}$P$_x$)$_2$ Superconductor*, Phys. Rev. B **81**, 184519 (2010).

30 M. M. Qazilbash, J. J. Hamlin, R. E. Baumbach, L. J. Zhang, D. J. Singh, M. B. Maple, and D. N. Basov, *Electronic Correlations in the Iron Pnictides*, Nature Phys. **5**, 647 (2009).

31 A. Kallel, H. Boller, and E. F. Bertaut, *Helimagnetism in Mnp-Type Compounds: Mnp, Fep, Cras, and CrAs$_{1-x}$Sb$_x$ Mixed Crystals*, J. Phys. Chem. Solid. **35**, 1139 (1974).

32 A. Yeh, Y.-A. Soh, J. Brooke, G. Aeppli, T. F. Rosenbaum, and S. M. Hayden, *Quantum Phase Transition in a Common Metal*, Nature **419**, 459 (2002).

33 R. Jaramillo, Y. Feng, J. Wang, and T. F. Rosenbaum, *Signatures of Quantum Criticality in Pure Cr at High Pressure*, PNAS **107**, 13631 (2010).

34 P. Pedrazzini and D. Jaccard, *The Critical Pressure of Chromium*, Physica B **403**, 1222 (2008).

35 J. G. Analytis, H.-H. Kuo, R. D. McDonald, M. Wartenbe, P. M. C. Rourke, N. E. Hussey, and I. R. Fisher, *Transport near a Quantum Critical Point in BaFe$_2$(As$_{1-x}$P$_x$)$_2$*, Nature Physics **10**, 194 (2014).

36 R. Yu, P. Goswami, Q. Si, P. Nikolic, and J. X. Zhu, *Superconductivity at the Border of Electron Localization and Itinerancy*, Nature Communications **4**, 2783 (2013).

37 T. Ito, H. Ido, and K. Motizuki, *Electronic Structure and Magnetic Properties in CrX (X=P, As, and Sb)*, J. Mag. Mag. Mater. **310**, e558 (2007).